\documentclass[twocolumn,english]{revtex4-1}
\usepackage[T1]{fontenc}
\usepackage{color}
\usepackage{amsmath}
\usepackage{amssymb}
\usepackage{graphicx}
\usepackage{esint}
\usepackage{lineno}
\usepackage{tabularx}

\definecolor{ablue}{rgb}{0.1,0.35,0.75}
\definecolor{agreen}{rgb}{0,0.6,0.4}

\usepackage[normalem]{ulem}

\makeatletter
\@ifundefined{textcolor}{}
{%
 \definecolor{BLACK}{gray}{0}
 \definecolor{WHITE}{gray}{1}
 \definecolor{RED}{rgb}{1,0,0}
 \definecolor{GREEN}{rgb}{0,1,0}
 \definecolor{BLUE}{rgb}{0,0,1}
 \definecolor{CYAN}{cmyk}{1,0,0,0}
 \definecolor{MAGENTA}{cmyk}{0,1,0,0}
 \definecolor{YELLOW}{cmyk}{0,0,1,0}
}

\renewcommand{\fnum@figure}{\textbf{Figure~\thefigure}}
\usepackage{units}
\usepackage{lipsum}

\usepackage{babel}

\makeatother

\usepackage{babel}
\begin{document}

\title{Intrinsic Lifetime of Higher Excitonic States in Tungsten Diselenide Monolayers}

\author{Samuel Brem$^1$}
\email{samuel.brem@chalmers.se}
\author{Jonas Zipfel$^2$}
\author{Malte Selig$^3$}
\author{Archana Raja$^4$}
\author{Lutz Waldecker$^5$}
\author{Jonas Ziegler$^2$}
\author{Takashi Taniguchi$^6$}
\author{Kenji Watanabe$^6$}
\author{Alexey Chernikov$^2$}
\author{Ermin Malic$^1$}
\affiliation{$^1$Chalmers University of Technology, Department of Physics, 41296 Gothenburg, Sweden}
\affiliation{$^2$University of Regensburg, Department of Physics, 93053 Regensburg, Germany}
\affiliation{$^3$Technical University Berlin, Institute of Theoretical Physics, 10623 Berlin, Germany}
\affiliation{$^4$Kavli Energy NanoScience Institute, University of California Berkeley,  Berkeley, USA}
\affiliation{$^5$Stanford University,  348 Via Pueblo Mall, Stanford, California 94305, USA}
\affiliation{$^6$National Institute for Materials Science, Tsukuba, Ibaraki 305-004, Japan}

\begin{abstract}
The reduced dielectric screening in atomically thin transition metal dichalcogenides allows to study the hydrogen-like series of higher exciton states in optical spectra even at room temperature. The width of excitonic peaks provides information about the radiative decay and phonon-assisted scattering channels limiting the lifetime of these quasi-particles. While linewidth studies so far have been limited to the exciton ground state, encapsulation with hBN has recently enabled quantitative measurements of the broadening of excited exciton resonances. Here, we present a joint experiment-theory study combining microscopic calculations with spectroscopic measurements on the intrinsic linewidth and lifetime of higher exciton states in hBN-encapsulated WSe$_2$ monolayers. Surprisingly, despite the increased number of scattering channels, we find both in theory and experiment that the linewidth of higher excitonic states is similar or even smaller compared to the ground state. Our microscopic calculations ascribe this behavior to a reduced exciton-phonon scattering efficiency for higher excitons due to spatially extended orbital functions.
\end{abstract}
\maketitle

Monolayer transition metal dichalcogenides (TMDs) show pronounced Coulomb phenomena \cite{ugeda2014giant,berghauser2014analytical, steinhoff2015efficient, wang2018colloquium, mueller2018exciton,baranowski2017probing,deilmann2018interlayer}, which in bulk materials become observable predominantly at very low temperatures. Electron-hole pairs in TMDs exhibit binding energies of up to \unit[0.5]{eV} giving rise to a Rydberg-like series of exciton states below the free particle bandgap \cite{chernikov2014exciton,hill2015observation,stier2018magnetooptics,robert2018optical}. While the relative position of excitonic resonances in optical spectra presents a fingerprint of Coulomb correlations \cite{steinhoff2014influence,raja2017coulomb, steinleitner2018dielectric}, their linewidth contains information about their coherence lifetime and the underlying many-particle scattering processes. Previous studies on exciton linewidths in TMDs \cite{singh2015intrinsic,palummo2015exciton,dey2016optical,christiansen2017phonon} have revealed efficient scattering into dark intervalley excitons \cite{selig2016excitonic} and have demonstrated strain-induced modifications of exciton-phonon scattering channels \cite{niehues2018strain}. However, these studies have been restricted to the 1s exciton ground state, since the higher order resonances were dominated by inhomogeneous broadening and were challenging to resolve and analyze at elevated temperatures. Furthermore, exciton-phonon scattering within the rich phase space of excited exciton states has not been theoretically investigated in TMDs yet. 
Recently, the encapsulation of TMD materials in the layered wide-bandgap insulator hexagonal boron nitride (hBN) has been shown to drastically reduce the inhomogeneous broadening of excitonic resonances \cite{stier2018magnetooptics,robert2018optical} and thereby enables to access temperature-dependent broadening of higher excitonic states.

\begin{figure}[t!]
\includegraphics[width=55mm]{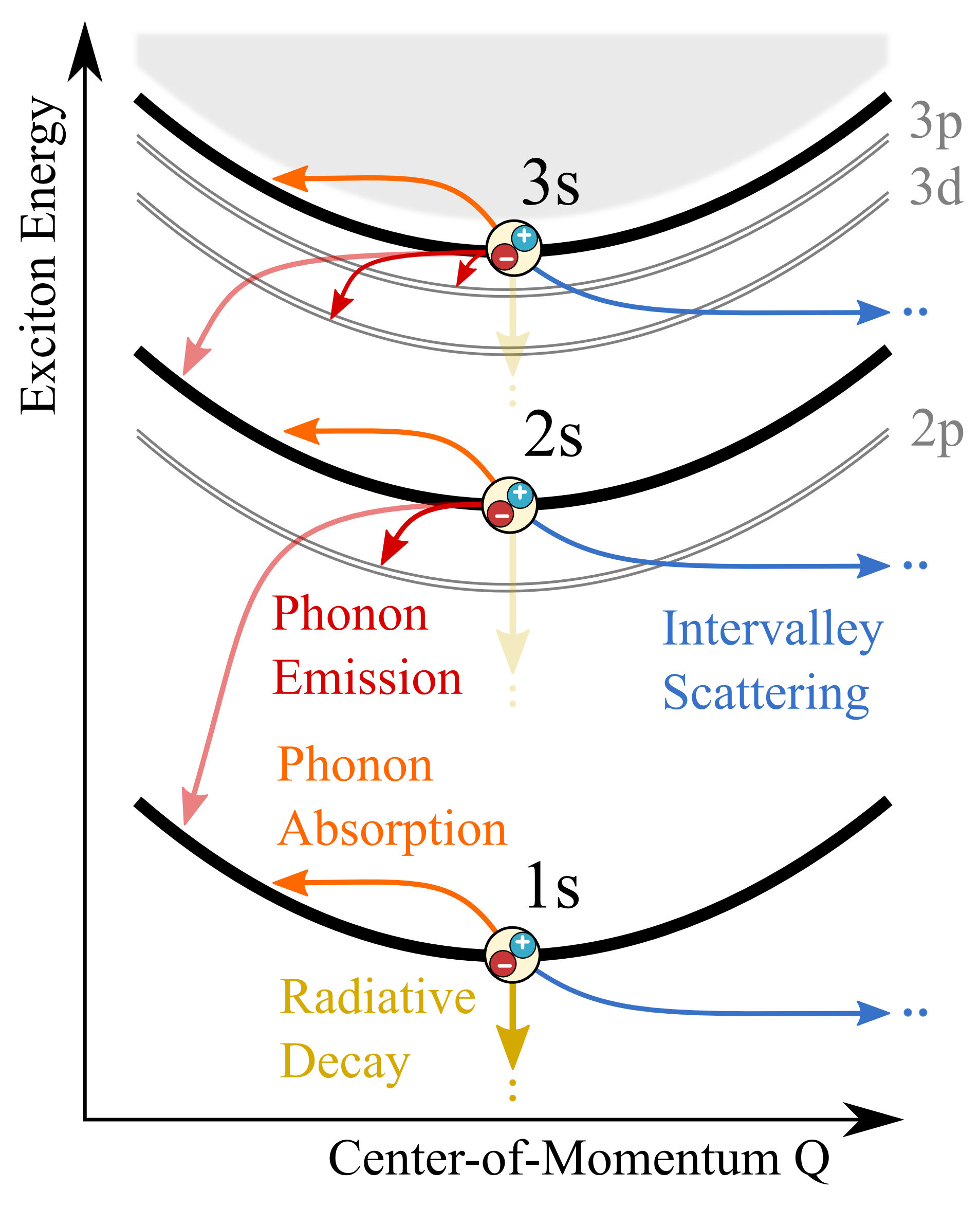}
\caption{Schematic illustration of the exciton bandstructure and possible scattering mechanisms. Optically generated excitons at zero center-of-mass momentum can decay either by radiative recombination (yellow arrow) or by scattering into a dark state with finite center-of-mass momentum. Apart from intravalley scattering via absorption of acoustic phonons (orange) or transitions into lower lying states after emission of a phonon (red), electron or hole can scatter into a different valley creating an intervalley exciton (blue).}
\label{fig:scheme} 
\end{figure}

In this work, we present a joint experiment-theory study on the homogeneous broadening of higher excitonic resonances and the underlying microscopic scattering mechanism for the exemplary case of hBN-encapsulated monolayer tungsten diselenide (WSe$_2$). Our calculations based on the density matrix formalism provide microscopic access to radiative decay and phonon-assisted exciton scattering channels across the full Rydberg-like series of excitonic states including intervalley excitons and symmetry forbidden dark states (e.g. p-type excitons), cf. Fig.\,\ref{fig:scheme}. 
The microscopic model is supported by reflectance contrast measurements on hBN-encapsulated WSe$_2$ monolayers, providing temperature-dependent linewidths of the three energetically lowest exciton resonances 1s, 2s and 3s.  
We reveal that although excited states exhibit a much larger phase space for possible relaxation channels, their linewidth stays comparable with the ground state and even decreases for states above 2s excitons. This reflects the increase of the exciton Bohr radii for excited states, which results in reduced excitonic form factors due to a smaller overlap of the initial and final state wavefunctions and thus a quenched exciton-phonon scattering efficiency. \\

\textbf{Theoretical approach.} To calculate the optical response of TMD monolayers, we use the Heisenberg equation of motion to derive the semiconductor Bloch equations \cite{koch2006semiconductor,kira2006many,haug2009quantum}. Here, we explicitly include the exciton-phonon interaction to account for non-radiative broadening of excitonic resonances.  Moreover, we consider the Coulomb interaction on a Hartree-Fock level, which yields an excitonic renormalization of interband transitions as well as  exciton-phonon transition energies. Within a second order Born-Markov approximation the linear optical absorption of a circularly($\sigma$)-polarized field reads \cite{kira2006many}
\begin{eqnarray} \label{eq:chi}
\alpha_\sigma(\omega)= \dfrac{e_0^2}{m_0^2 \epsilon_0 n c_0 \omega}\lvert M_\sigma\rvert^2\sum_\nu \Im\text{m}\big(\dfrac{\lvert\Phi_\nu(r=0)\rvert^2}{E_\nu-\hbar\omega-i\Gamma_\nu}\big).
\end{eqnarray}
Here, we used the $p \cdot A$ gauge, so that the overall strength of the response is determined by the projection of the interband momentum matrix element on the light polarization  $M_\sigma$, which is determined analytically within a two band $k\cdot p$ approach \cite{xiao2012coupled}, cf. supporting information (SI). The constants $e_0$ and $m_0$ denote the elementary charge and the electron rest mass, while $c_0$ and $n$ are the vacuum light velocity and refractive index, respectively. Furthermore, the optical response is given by a sum of Lorentzian peaks, whose position and oscillator strength is determined by the eigen energy $E_\nu$ and wavefunctions $\Phi_\nu$ of exciton states $\nu$.  The latter are obtained by solving the Wannier equation \cite{koch2006semiconductor,kira2006many,selig2018dark} within an effective mass approximation and a thin-film approximation of the Coulomb potential to account for a non-uniform dielectric environment \cite{rytova1967ns, keldysh1979lv}. Details about the solution of the Wannier equation and the used ab initio parameters \cite{kormanyos2015k,laturia2018dielectric} are given in the SI.

The decay rate $\Gamma_\nu=\Gamma^\text{phon}_\nu+\Gamma^\text{rad}_\nu$ determines the broadening of excitonic resonances in an absorption spectrum. It corresponds to the inverse lifetime of coherent excitons, i.e. optically generated excitons at zero center-of-mass momentum. They can decay either by recombination $\Gamma^\text{rad}_\nu$ (radiative dephasing) or scattering with phonons $\Gamma^\text{phon}_\nu$ into a state with finite center-of-mass momentum. Taking into account a self-consistent coupling between the light field and the induced microscopic polarization \cite{kira2006many,knorr1996theory}, we obtain the radiative broadening 
\begin{eqnarray} \label{eq:Gamma_rad}
\Gamma^\text{rad}_\nu=\dfrac{\hbar e_0^2}{2 m_0^2\epsilon_0 n c_0} \lvert M_\sigma\rvert^2 \dfrac{\lvert\Phi_\nu(r=0)\rvert^2}{E_\nu},
\end{eqnarray}
The non-radiative dephasing due to exciton-phonon-scattering reads \cite{thranhardt2000quantum,selig2018dark,brem2018exciton} 
\begin{eqnarray} \label{eq:Gamma_phon}
\Gamma^\text{phon}_\nu=\pi \sum_{\pm,\lambda,\mu,\mathbf{q}}  \lvert G^{\nu \mu}_{\lambda\mathbf{q}}\rvert^2 \hat{n}^{\pm}_{\lambda\mathbf{q}}  \delta(E^{\mu}_{\mathbf{q}}-E^\nu_0 \pm \hbar {\omega}_{\lambda\mathbf{q}}).
\end{eqnarray} 
Here, the occupation factor $\hat{n}^{\pm}_{\lambda\mathbf{q}}=\frac{1}{2}\pm\frac{1}{2}+n_{\lambda\mathbf{q}}$ accounts for the number of phonons $n_{\lambda\mathbf{q}}$ in the mode $\lambda$ and the momentum $\mathbf{q}$ weighting the emission ($+$) and absorption processes ($-$). In the applied Markov approximation we require strict energy and momentum conservation for the scattering from  the lowest energy state in the light cone $E^\nu_{q=0}$ to an exciton $E^\mu_\mathbf{q}$ with a non-zero center-of-mass momentum $q$ under interaction with a phonon with the energy $\hbar{\omega}_{\lambda\mathbf{q}}$.
The scattering cross-section for exciton transitions $(\nu,0)\rightarrow(\mu,\mathbf{q})$ is determined by the overlap of the initial and final state exciton wavefunctions in Fourier space $\tilde{\Phi}$ shifted by the scattering momentum and reads
\begin{eqnarray} \label{eq:ex-ph}
G^{\nu \mu}_{\lambda\mathbf{q}}=\sum_{\mathbf{k},\alpha=e,h} g^{\alpha\mathbf{k}}_{\lambda \mathbf{q}}\,\tilde{\Phi}_{\nu}^{\ast}(\mathbf{k}) \tilde{\Phi}_\mu(\mathbf{k+q_\alpha}).
\end{eqnarray}
The relative momentum transferred to the electron (hole) constituent when the entire exciton gains a center-of-mass momentum $q$ is denoted by $q_{e(h)}$ with $q=q_e+q_h$. The exciton index $\nu$ here acts as a compound index containing principal quantum number, angular momentum, electron/hole valley and spin configuration. For the carrier-phonon coupling $g$ we use the deformation potential approximations for acoustic and optical phonons deduced from density functional perturbation theory (DFPT) in Ref. \cite{jin2014intrinsic}. Here, the electron-phonon coupling in the vicinity of high-symmetry points only depends on the transferred momentum $q$, so that $G^{\nu \mu}_{\lambda\mathbf{q}} \propto \mathcal{F}_{\nu\mu}(\mathbf{q_\alpha})= \int d^2r \Phi_\nu^\ast(\mathbf{r})e^{i\mathbf{q_\alpha r}}\Phi_\mu(\mathbf{r})
$. The excitonic form factor $\mathcal{F}_{\nu\mu}$ accounts for the geometry of the exciton wavefunctions involved in the scattering process, which will be crucial for the interpretation of our results.

 Throughout this work, we only consider the A-exciton series and thus focus on the spin configuration composing the lowest optically allowed transition at the K point. However we take into account phonon scattering into all minima (maxima) of the conduction (valence) band with the same spin configuration including intervalley scattering of electrons to the $\Lambda,\Lambda'$ and $K'$ valley and hole scattering to the $\Gamma$ or $K'$ point. Furthermore, we include the full excitonic Rydberg series for intra- as well as intervalley excitons reachable via phonon-assisted transitions, cf. Fig.\,\ref{fig:scheme}. Further details about the used matrix-elements and the calculation of scattering amplitudes are given in the SI.
 \\

\begin{figure}[!t]
\includegraphics[width=\linewidth]{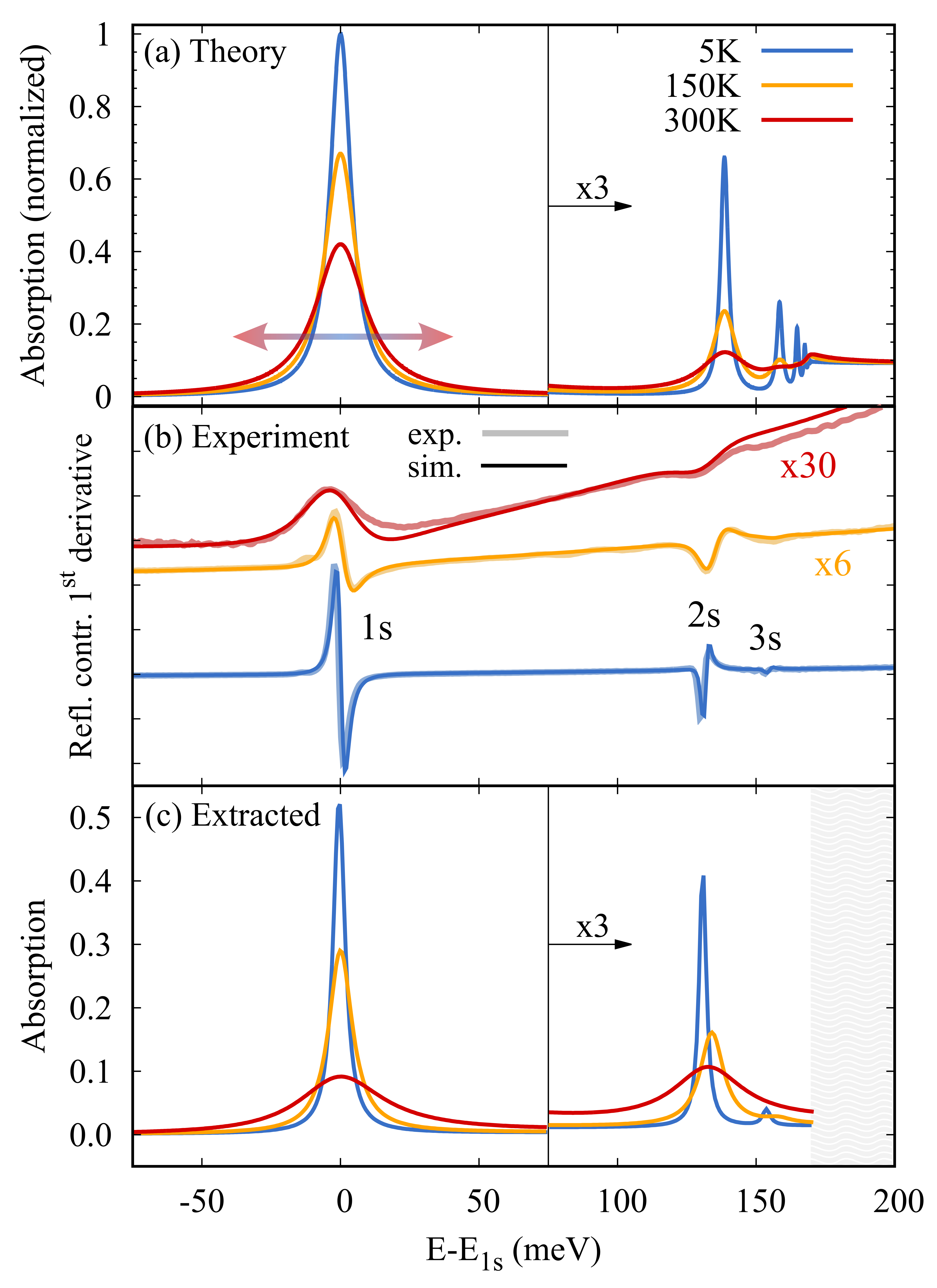}
\caption{(a) Theoretically calculated absorption spectra of an hBN-encapsulated WSe$_2$ monolayer at different temperatures. Due to the strong Coulomb interaction a clear hydrogen-like series of bound exciton resonances appears below the single particle bandgap. At low temperatures the intrinsic broadening allows to clearly resolve several excited states in the spectrum and the first excited state is visible even at $\unit[300]{K}$. (b) Experimentally measured reflectance contrast derivatives of an hBN-encapsulated WSe$_2$ monolayer together with the simulated curves from the multi-Lorentzian model. The data are vertically offset and rescaled for clarity. (c) The corresponding extracted experimental absorption spectra of the 1s, 2s, and 3s exciton resonances.
}
\label{fig:spectra} 
\end{figure}

\textbf{Exciton spectroscopy.} Linear reflectance measurements are performed on a WS$_2$ monolayer sample encapsulated between ultra-thin layers of hexagonal boron nitride and placed on a SiO$2$/Si substrate. The spectra are acquired at lattice temperatures from 5\, to 300\,K using a spectrally broad whitelight source for illumination. Spectrally-resolved reflectance signals are recorded both on the sample and on the SiO$2$/Si substrate reference. The data are presented in terms of reflectance contrast, defined by the relative change in reflectance on the sample with respect to the reference. For quantitative analysis, the optical response is parameterized by a multi-Lorentzian dielectric function, with the exciton 1s, 2s, and 3s states each represented by a Lorentzian peak resonance. The individual peak parameters are adjusted to match the measured reflectance contrast derivatives taking into account multi-layer interference effects in the studied structure using a transfer-matrix formalism. This approach allows for a quantitative analysis of the exciton peak parameters, such as transition energies and linewidths, as well as their contributions to the optical absorption of the monolayer. Additional details of the experimental procedure and data analysis are provided in the SI.

\textbf{Results.} The numerical evaluation of Eq. (\ref{eq:chi}) provides microscopic access to the excitonic absorption spectrum of an arbitrary TMD material. Figure \ref{fig:spectra}(a) shows the calculated spectra of an hBN-encapsulated WSe$_2$ monolayer at $5$, $150$ and $\unit[300]{K}$, presented as function of energy relative to the exciton ground state resonance. 
At energies larger than the band gap we observe an absorption continuum reflecting the excitation of unbound electrons and holes. Furthermore, we obtain a hydrogen-like series of exciton resonances at their respective binding energies below the free particle bandgap. Apart from the strong absorption from the $1\text{s}$ ground state, we also observe higher excitonic resonances $2\text{s}, 3\text{s},...$ with decreasing oscillator strength. We find about one order of magnitude difference between the oscillator strength of the  $1\text{s}$ and the $2\text{s}$ state. With increasing temperatures, the absorption lines become broader due to an enhanced exciton-phonon interaction and the higher order resonances start to spectrally overlap. Nevertheless, over a range of temperatures the intrinsic broadening allows us to resolve several excited states in the spectrum. Importantly, the $2s$ resonance is clearly visible even at room temperature.

Experimentally measured data from hBN-encapsulated WSe$_2$ monolayer are presented in Fig.\,\ref{fig:spectra}\,(b) as reflectance contrast derivatives together with the results of the multi-Lorentzian simulation and in (c) as the corresponding extracted experimental absorption spectra at $5$, $150$ and $\unit[300]{K}$. The absorption is shown only in the range of the three exciton states appearing in the measured spectra and considered for the data analysis. The exciton 1s and 2s states are clearly observed at all temperatures and the 3s state is detected for temperatures below $\unit[150]{K}$. As the temperature increases, the exciton peaks broaden under conservation of both peak areas and relative energy positions within a few meV. All changes are continuous, as further illustrated by the full temperature series presented in the SI.

\begin{figure*}[!t]
\includegraphics[width=150mm]{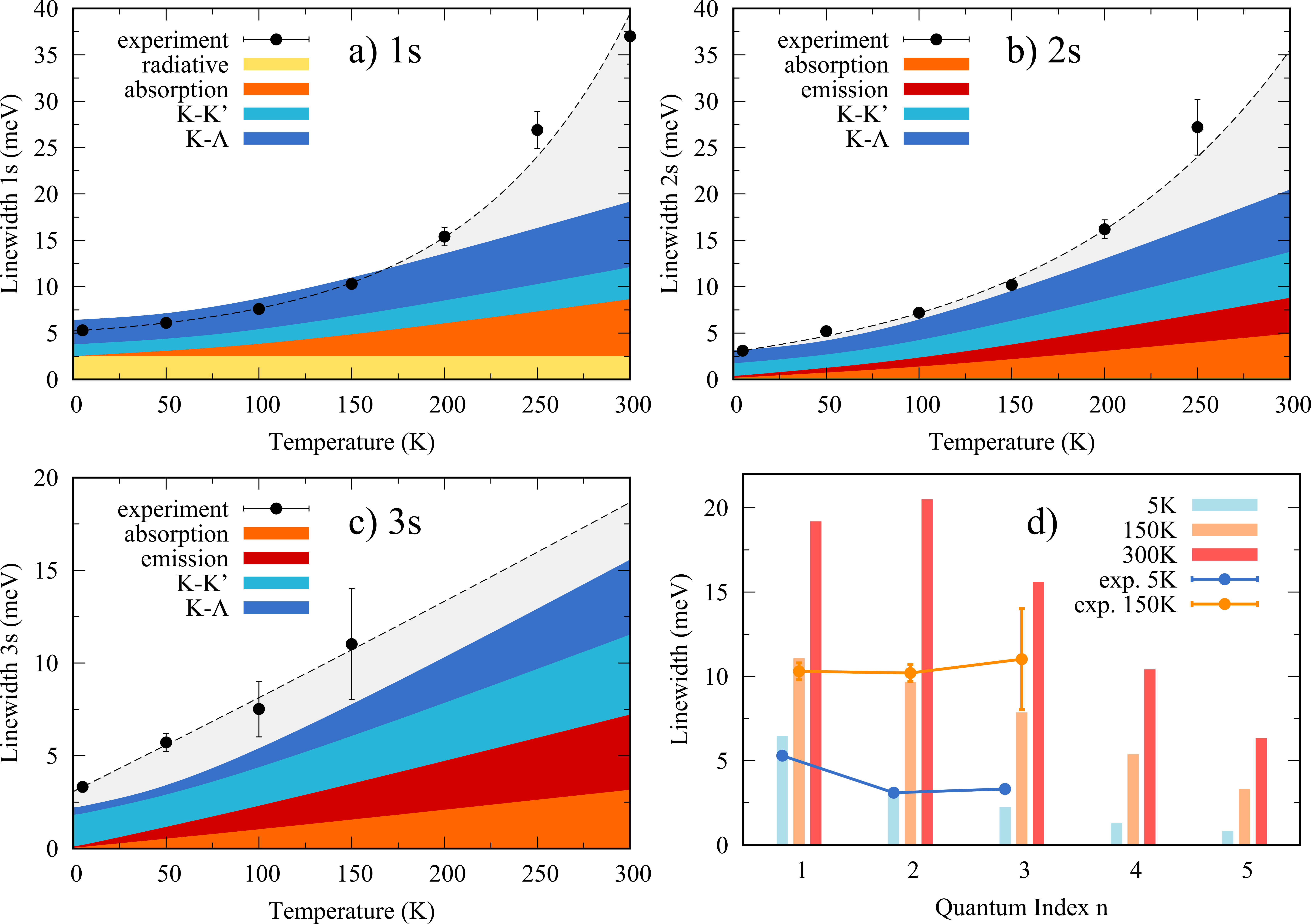}
\caption{Temperature-dependent linewidths of the (a) 1s, (b) 2s, and (c) 3s A exciton. The black points with error bars show the linewidth extracted from reflectance contrast measurements (dashed line shows a guide-to-the-eye using an exponential fit function). The theoretically calculated homogeneous linewidths are decomposed and color-coded in contributions of radiative decay, intravalley absorption/emission and intervalley scattering into the indirect K-$\Lambda$ and K-K' exciton, cf. Fig.\,\ref{fig:scheme}. (d) The linewidth of the bright $n$s excitons as function of $n$ for different temperatures. The microscopic model yields a decreasing trend for high $n$ resulting from a reduced phonon scattering efficiency for excited states.
}
\label{fig:linewidth} 
\end{figure*}

In the following, we analyze the temperature dependence of the homogeneous total linewidth in the absorption, defined as full-width-at-half-maximum, and discuss the underlying microscopic scattering mechanisms. Figure \ref{fig:linewidth} shows the linewidth as a function of the lattice temperature for the (a) 1s, (b) 2s and (c) 3s states. The black points show the linewidth extracted from reflectance contrast measurements with an exponential dashed line included as guide to the eye. Theoretical results are presented as a color-coded decomposition illustrating the individual contributions from different scattering channels. Specifically, these include radiative broadening (yellow), phonon absorption (orange), phonon emission (red) and intervalley scattering processes (light and dark blue), as illustrated in Fig.\,\ref{fig:scheme}. 

The broadening of the 1s ground state is determined by the three channels depicted in Fig.\,\ref{fig:scheme}: (i) A temperature independent broadening of about $\unit[2.5]{meV}$ results from the coherent radiative decay of the excitonic polarization (yellow area in Fig.\,\ref{fig:linewidth}(a)). (ii) Since $1\text{s}$ is the energetically lowest spin-like state at the K point, the only efficient intravalley phonon scattering process is the absorption of long range, small momentum acoustic phonons giving rise to a linear increase of the linewidth with temperature (orange area). (iii) Finally, there is also scattering towards intervalley exciton states. 
These include, in particular the $K-\Lambda$ excitons (hole located at the $K$ and electron at the $\Lambda$ valley), which due to the two times higher electron mass at the $\Lambda$ valley have a larger binding energy than the direct $K-K$ exciton. In tungsten-based TMDs the electronic band splitting between $K$ and $\Lambda$ point turns out to be smaller than the difference in exciton binding energies, so that the dark $K-\Lambda$ state is energetically below the bright state. In addition, the valley dependent spin-orbit splitting of the conduction band also gives rise to lower lying $K-K'$ excitons in tungsten-based TMDs. Therefore, intervalley scattering via emission of phonons is very efficient in tungsten based TMDs and provides a significant contribution to the linewidth even at $\unit[0]{K}$ (light and dark blue area in Fig.\,\ref{fig:linewidth}(a)) \cite{selig2016excitonic}.  For the investigated, prototypical case of hBN-encapsulated WSe$_2$, we find an additional broadening of $\unit[3.9]{meV}$ at $\unit[0]{K}$ due to inter-valley relaxation. 

When comparing the excited states with the ground state we find that, although the underlying scattering mechanisms include additional processes, the total broadening of $5$ to $\unit[20]{meV}$ is interestingly comparable for all three states. In addition to the mentioned scattering channels, the $2\text{s}$ and $3\text{s}$ excitons can efficiently scatter to lower lying states within the same valley via emission of phonons (cf. red arrows in Fig.\,\ref{fig:scheme}). Furthermore, there are more intervalley scattering channels available for excited states, since here transitions to e.g. $K-\Lambda$ 2p states become possible. Therefore, these processes contribute to additional broadening of the exciton resonances with increasing quantum number $n$. However, we observe both in theory and experiment a similar broadening for $1\text{s}$, $2\text{s}$ and $3\text{s}$ and even a decreased linewidth for the excited states at low temperatures.

To understand this somewhat counter-intuitive result we have to consider the influence of the exciton wavefunctions on radiative and non-radiative transition probabilities. With increasing principal quantum index $n$, exciton orbital functions become larger in space, which decreases the radiative recombination efficiency $\propto \lvert\Phi_\nu(r=0)\lvert^2$, cf. Eq. \eqref{eq:Gamma_rad}. This explains the strong reduction of the radiative broadening from about $\unit[2.5]{meV}$ for 1s to about $\unit[0.2]{meV}$ for 2s (yellow area in Fig.\,\ref{fig:linewidth}).
Moreover, the spatially larger orbitals of excited states correspond to narrower wavefunctions in Fourier (momentum) space. Since the phonon-scattering efficiency is given by the overlap of initial and final state wavefunctions in momentum space, cf. Eq. \eqref{eq:ex-ph}, larger exciton radii lead to smaller scattering probabilities reflecting the reduced quantum mechanical momentum uncertainty \cite{toyozawa1958theory}. Furthermore, the spatial oscillation of the wavefunctions of excited states additionally quenches the overlap integral in Eq. \eqref{eq:ex-ph}. For the absorption of long range acoustic phonons via $n\text{s}\rightarrow n\text{s}$ (orange contributions in Fig.\,\ref{fig:linewidth}), the reduced scattering efficiency can be shown analytically by considering the excitonic form factor. Due to the small momentum transfer $q_0$ for these processes, it holds $\Gamma_{\nu\rightarrow\nu}\propto \langle\nu\lvert e^{iq_0r}\rvert\nu\rangle \approx 1-\frac{1}{2}q_0^2\langle \nu\lvert r^2 \rvert \nu\rangle$. Since $\langle \nu\lvert r^2 \rvert \nu\rangle$ is a measure of the spatial variance of the probability function $\lvert \Phi_\nu(\mathbf{r})\rvert^2$, we find a decreased efficiency for the phonon absorption due to the increased orbital size of 2s and 3s excitons.

The additional phonon emission channels into 2p states of $K-K$ and $K-\Lambda$ excitons  compensate the loss of oscillator strength and scattering efficiency for the 2s state, cf. Fig.\,\ref{fig:linewidth}\,(b), yielding a comparable room temperature broadening of about $\unit[20]{meV}$ for both 1s and 2s states. However, the above discussed quenching of individual transition probabilities already gives rise to narrower 3s lines (Fig.\,\ref{fig:linewidth}\,(c)) despite an even large phase space of lower lying final states and associated emission relaxation channels. In particular, scattering into the $K'$-valley becomes more dominant, since the 3s state lies energetically above the free particle continuum of the $K-K'$ exciton providing a large phase space of final quasi-resonant states.

Finally, in Fig.\,\ref{fig:linewidth}\,(d) we show the homogeneous linewidth of  exciton resonances $n\text{s}$ as a function of their principal quantum number $n$ for three different temperatures. As discussed above, the increased phase space leads to a slight increase of the linewidth from 1s to 2s at room temperature. However, for larger quantum numbers or at lower temperatures we observe a clear monotonous decrease of the broadening with $n$, reflecting the strongly decreasing scattering cross section due to the larger exciton Bohr radii. We predict significantly narrower linewidths of 4s and 5s states compared to the 1s ground state. Our prediction is similar to the findings of the experimental and theoretical studies on Rydberg excitons in copper oxide \cite{kazimierczuk2014giant,stolz2018interaction}, where a decrease of the linewidth of $1/n^3$ with the principal quantum number $n$ was found. 

For the three lowest exciton states, we obtain a good agreement between experimentally observed and theoretically predicted linewidths for temperatures of up to $\unit[150]{K}$.
To reproduce the strong increase in the linewidth up to room temperature (grey-shaded area), one needs to take into account non-Markovian effects such as off-resonant interaction with phonons beyond energy conservations \cite{selig2016excitonic}, as well as the appearance of phonon sidebands \cite{christiansen2017phonon}. These effects, in addition to the here presented Markovian contributions give rise to a super linear increase of the full linewidth at elevated temperatures. In particular, the emergence of asymmetric phonon side bands can clearly be seen in the deviation of the  measured reflectance spectra from the simulated symmetric Lorentzians around the 1s state at $\unit[300]{K}$, cf. Fig. \ref{fig:spectra}(b). As reported in Ref.\cite{christiansen2017phonon} phonon-assisted optical transitions significantly influence the exciton lineshape in WSe$_2$ for temperatures above $\unit[150]{K}$, since here the energetically lower lying dark states give rise to side bands below and above the exciton main resonance. A homogeneous width of the exciton main line of $\unit[26]{meV}$ has been reported for WSe$_2$ monolayers on SiO$_2$ at room temperature, while the phonon sidebands have been shown to yield an additional asymmetric broadening with a total FWHM of $\unit[45]{meV}$ \cite{christiansen2017phonon}. The resulting ratio between FWHM and homogeneous linewidth of about $0.58$ corresponds well to the ratio of $0.51$ between the here calculated $\unit[19]{meV}$ homogeneous linewidth and the measured $\unit[37]{meV}$. The reduced homogeneous linewidth in our work is a result of the strongly increased dielectric screening from the hBN encapsulation, which predominantly modifies the exciton-phonon scattering channels via larger exciton orbitals and different intervalley separations.

Finally, in this study we have focused on the intrinsic broadening mechanisms in a homogeneous system, i.e. assuming a perfect lattice periodicity and a spatially constant dielectric background. While these assumptions have turned out to widely reproduce the linewidth in exfoliated, hBN-encapsulated TMDs, effects resulting from inhomogeneities will have a much larger impact in non-encapsulated samples, e.g. as-exfoliated flakes on SiO$_2$ substrates. Here both, the spatial fluctuation of resonance energies due to dielectric inhomogeneities as well as the possible elastic scattering of excitons at lattice defects should contribute to the broadening of excited and ground state. The scattering with defects, in particular, is expected to stronger influence the lifetime of excited states in contrast to the ground state, due to a smaller number of resonant final states for the later, necessary for elastic scattering. 
Similarly, spatial variations of the band gap due to dielectric disorder should also affect the resonance energy of the excited states by a larger degree due to reduced cancellation effects of the bandgap renormalization and binding energy \cite{cho2018environmentally}, yielding an increasing inhomogeneous broadening with the quantum number $n$. Therefore, the small constant offset between the calculated intrinsic linewidth and the measured broadening of the 3s state of a few meV (Fig.\,\ref{fig:linewidth}\,(c)) can be assigned to the presence of residual inhomogeneities potentially remaining in the hBN-encapsulated samples.\\

\textbf{Conclusion.} 
We have presented a joint experiment-theory study addressing the microscopic mechanisms governing the intrinsic broadening of higher exciton states in atomically thin WSe$_2$ monolayers. We reveal both in experiment and theory that the higher excitonic states surprisingly show either a similar or even smaller linewidth compared to the exciton ground state - despite the larger number of final states for scattering events. Importantly, while the small Bohr radius of the ground state enables both strong light-matter coupling and efficient absorption of acoustic phonons, these processes become much weaker for excited states. The spatially extended orbital functions of excited states give rise to lower oscillator strength and reduced exciton-phonon scattering efficiency, resulting from smaller overlaps of initial and final state wavefunctions in momentum space. On the other hand, transitions into energetically lower lying $p$ exciton states lead to an additional broadening mechanism resulting in comparable linewidths for the ground and excited exciton states. The gained insights should strongly contribute to the fundamental understanding of the exciton physics and carrier-lattice interactions in atomically thin transition metal dichalcogenide and guide future studies. \\

\textbf{Acknowledgements}
The Chalmers group acknowledges financial support from the European Unions Horizon 2020 research and innovation program under grant agreement No 696656 (Graphene Flagship) as well as from the Swedish Research Council (VR). M.S. acknowledges the support from  the Deutsche Forschungsgemeinschaft (DFG) through SFB 951 and the School of Nanophotonics (SFB 787).
A.C. and J.Z. acknowledge financial support by the DFG via Emmy Noether Grant CH 1672/1-1 and Collaborative Research Center SFB 1277 (B05).
A.R. gratefully acknowledges funding through the Heising-Simons Junior Fellowship within the Kavli Energy NanoScience Institute at the University of California, Berkeley.
L.W. acknowledges support by the Alexander von Humboldt foundation. A.R. and L.W. acknowledge support by the Gordon and Betty Moore Foundation's EPiQS program through grant GBMF4545. Growth of hexagonal boron nitride crystals was supported by the Elemental Strategy Initiative conducted by the MEXT, Japan and the CREST (JPMJCR15F3), JST.

\onecolumngrid
\appendix
\newpage
\noindent \begin{center}
\textbf{\large Intrinsic Lifetime of Higher Excitonic States in Tungsten Diselenide Monolayers   \\ 
\vspace*{5mm}
--SUPPLEMENTARY MATERIAL--}
\par\end{center}

\begin{center}
Samuel Brem$^{1*}$, Jonas Zipfel$^2$, Malte Selig$^3$, Archana Raja$^4$, Lutz Waldecker$^5$, Jonas Ziegler$^2$, Takashi Taniguchi$^6$, Kenji Watanabe$^6$, Alexey Chernikov$^2$, Ermin Malic$^1$  {\small }\\
{\small $^{1}$ }\textit{\small Chalmers University of Technology, Department of Physics, 41296 Gothenburg, Sweden}\\
{\small $^{2}$ }\textit{\small University of Regensburg, Department of Physics, 93053 Regensburg, Germany}\\
{\small $^{3}$ }\textit{\small Technical University Berlin, Institute of Theoretical Physics, 10623 Berlin, Germany}\\
{\small $^{4}$ }\textit{\small Kavli Energy NanoScience Institute, University of California Berkeley,  Berkeley, USA}\\
{\small $^{5}$ }\textit{\small Stanford University,  348 Via Pueblo Mall, Stanford, California 94305, USA}\\
{\small $^{6}$ }\textit{\small National Institute for Materials Science, Tsukuba, Ibaraki 305-004, Japan}\\
\par\end{center}{\small \par}

\section{Hamilton Operator and Material Parameters}

The properties of the TMD monolayer are described by the following many-particle Hamiltonian:

\begin{eqnarray} \label{eq:hamilton}
H&=&H_0+H_{\text{Coul}}+H_{\text{el-l}}+H_{\text{el-ph}} \\
 &=& \sum_{\alpha \mathbf{k}} \varepsilon_{\alpha\mathbf{k}} a^\dagger_{\alpha\mathbf{k}}a^{\phantom\dagger}_{\alpha\mathbf{k}} + \sum_{\lambda \mathbf{q}} \hbar \omega_{\lambda\mathbf{q}} b^\dagger_{\lambda\mathbf{q}}b^{\phantom\dagger}_{\lambda\mathbf{q}} \nonumber\\
&\ \ \ +& \dfrac{1}{2}\sum_{\alpha \beta \mathbf{k}\mathbf{k}'\mathbf{q}} W_\mathbf{q} a^\dagger_{\alpha\mathbf{k+q}}a^\dagger_{\beta\mathbf{k'-q}}a^{\phantom\dagger}_{\beta\mathbf{k}'}a^{\phantom\dagger}_{\alpha\mathbf{k}}\nonumber\\
 &\ \ \ +& \dfrac{e_0}{m_0}\sum_{\alpha \beta \mathbf{k}} \mathbf{M}^{\alpha\beta}_\mathbf{k}\cdot \mathbf{A} \mathbf{} a^\dagger_{\alpha\mathbf{k}}a^{\phantom\dagger}_{\beta\mathbf{k}}\nonumber\\
&\ \ \ +& \sum_{\alpha \lambda \mathbf{k}\mathbf{q}} g^{\alpha\mathbf{k}}_{\lambda \mathbf{q}} a^\dagger_{\alpha\mathbf{k+q}}a^{\phantom\dagger}_{\alpha\mathbf{k}}(b^{\phantom\dagger}_{\lambda\mathbf{q}}+b^{\dagger}_{\lambda,-\mathbf{q}})
\end{eqnarray}

Here  $a^{(\dagger)}_{\alpha\mathbf{k}}$ denotes the annihilation (creation) operator of an electron in band $\alpha=c,v$ with momentum $\mathbf{k}$, and $b^{(\dagger)}_{\lambda\mathbf{q}}$ annihilates (creates) a phonon in mode $\lambda$ with momentum $\mathbf{q}$. For the electronic bandstructure $\varepsilon_{\alpha\mathbf{k}}$ we use the effective mass approximations deduced from ab initio calculation (PBE) in ref. \cite{kormanyos2015k}, while the phonon dispersion $\omega_{\lambda\mathbf{q}}$ is descibed in Debye (long range acoustic) or Einstein approximation (optical and short range acoustic) with sound velocities and energies adopted from DFPT calculations in ref.\cite{jin2014intrinsic}. For the Coulomb interaction $W_q$ we derive a modified form of the potential in ref.'s \cite{rytova1967ns,keldysh1979lv} for charges in a thin film of thickness d surrounded by a dielectric environment. In this work we explicitly take into account anisotropic dielectric tensors. Solving the Poisson equation with the above described boundary conditions yields $W_q=V_q/\epsilon_{scr}(q)$, with the bare 2D-Fourier transformed Coulomb potental $V_q$ and a non-local screening,

\begin{eqnarray} \label{eq:keldish}
\epsilon_{scr}(q)= \kappa_1 \tanh(\dfrac{1}{2}[\alpha_1dq-\ln(\dfrac{\kappa_1-\kappa_2}{\kappa_1+\kappa_2})]),
\end{eqnarray}

where $\kappa_{i}=\sqrt{ \epsilon^{\parallel}_i \epsilon^{\bot}_i}$ and $\alpha_i=\sqrt{ \epsilon^{\parallel}_i /\epsilon^{\bot}_i}$ account for the parallel and perpendicular component of the dielectric tensor $\epsilon$ of the monolayer ($i=1$) and the environment ($i=2$).
The momentum matrix element $\mathbf{M}^{\alpha\beta}_\mathbf{k}=-i\hbar \langle \alpha \mathbf{k}|\triangledown|\beta\mathbf{k}\rangle$ is derived from a two band $k\cdot p$ Hamiltonian, which in vicinity of the K point yields \cite{xiao2012coupled}

\begin{eqnarray} \label{eq:optical}
|\mathbf{M}^{vc}_\mathbf{k}\cdot \mathbf{e}_\sigma|^2\equiv|M_\sigma|^2=\dfrac{1}{2}[\dfrac{a_0m_0t}{\hbar}(1+\sigma)]^2.
\end{eqnarray}

The next neighbor hopping integral $t=\hbar/a_0 \sqrt{E_g/(m_e+m_h)}$ is determined by the effective masses $m_{e/h}$ of electrons and holes and the single particle bandgap $E_g$ at the K-point, while $\sigma=\pm 1$ for left-(right-)handed circularly polarized light.
In Table \ref{table:parameter} we summarize the used parameters for hBN encapsulated WSe$_2$ for the evaluation of Eq. \eqref{eq:keldish} and \eqref{eq:optical}.
\begin{table}[b] 
\centering
\begin{tabularx}{\textwidth}{X X X X X}
\hline
\hline
Parameter&& Value && Reference \\
\hline
Lattice constant $a_0$ && \unit[0.334]{nm} && \cite{laturia2018dielectric} \\
Layer thickness $d$ && \unit[0.652]{nm} && \cite{laturia2018dielectric} \\
dielec. para. $\epsilon^{\parallel}_{WSe2}$ && 15.1 && \cite{laturia2018dielectric} \\
dielec. perp. $\epsilon^{\bot}_{WSe2}$ && 7.5 && \cite{laturia2018dielectric} \\
dielectr. hBN $\kappa_{hBN}$ && 4.5 && \cite{geick1966normal} \\
Bandgap $E_g$(hBN enc.) && \unit[2.38]{eV}(\unit[2.02]{eV}) && \cite{kormanyos2015k,cho2018environmentally} \\
Electron mass (at K) $m_e$ && 0.29$m_0$ && \cite{kormanyos2015k} \\
Hole mass (at K) $m_h$ && 0.36$m_0$ && \cite{kormanyos2015k} \\
\hline
\end{tabularx}
\caption{List of the used parameters for the calculation of exciton eigenenergies and wavefunctions in hBN encapsulated WSe$_2$ monloayers.
\label{table:parameter} 
}\end{table}

Finally the electron phonon coupling $g^{\alpha\mathbf{k}}_{\lambda \mathbf{q}}$ is approximated with the generic form of a deformation potential

\begin{eqnarray} \label{eq:el-ph}
g^{\alpha\mathbf{k}}_{\lambda \mathbf{q}}\approx \sqrt{\dfrac{\hbar}{2\rho A \omega_{\lambda \mathbf{q}}}} D^{\alpha}_{\lambda \mathbf{q}}.
\end{eqnarray}

Here $\rho$ denotes the surface mass density of the monolayer and $A$ the area of the system. For the coupling constant $D^{\alpha}_{\lambda \mathbf{q}}$  we adopt the approximations deduced from DFPT calculations in ref. \cite{jin2014intrinsic}, where long range acoustic phonons couple linear in momentum $D^{\alpha}_{\lambda \mathbf{q}}\rvert_{\text{intra ac}}\approx D^{\alpha (1)}_\lambda q$, while optical phonons and short range acoustic modes couple with a constant strength $D^{\alpha}_{\lambda \mathbf{q}}\rvert_{\text{inter ac,opt}}\approx D^{\alpha (0)}_\lambda$ in vicinity of high symmetry points. We take into account the LA,TA,LO,TO and A1 mode for intravalley as well as scattering of electrons to the $\Lambda,\Lambda'$ and $K'$ valley and hole scattering to the $\Gamma$ or $K'$ point. The constants $D^{(0)}$ and  $D^{(1)}$ for all possible intra and intervalley scattering channels are listed in ref. \cite{jin2014intrinsic}.

\section{Wannier Equation and Intervalley Scattering}

The linear optical response of a system is obtained from the Heisenberg equation of motion for the microscopic polarisation $p_{\mathbf{k}\mathbf{Q}}=\langle a^\dagger_{c,\mathbf{k}+\alpha\mathbf{Q}}a^{\phantom\dagger}_{v,\mathbf{k}-\beta\mathbf{Q}}\rangle$ \cite{kira2006many}. We use relative (k) and center-of mass coordinates (Q) with $\alpha(\beta)=m_{c(v)}/(m_c+m_v)$. The effective Coulomb interaction between charge carriers in TMDs leads to a strong coupling of polarisations at different relative momenta k, yielding an excitonic eigen spectrum for interband transitions. To decouple the equations of motion we perform a basis transformation by expanding the polarisation in terms of exciton wave functions $p_{\mathbf{k}\mathbf{Q}}=\sum_\nu \tilde{\Phi}_{\nu\mathbf{Q}}(\mathbf{k}) P_{\nu\mathbf{Q}}$. To diagonalize the equations of motion for $P_{\nu\mathbf{Q}}$, the basis functions have to fullfill the Wannier equation,
\begin{eqnarray} \label{eq:wannier}
 (\varepsilon_{c,\mathbf{k}+\alpha\mathbf{Q}}-\varepsilon_{v,\mathbf{k}-\beta\mathbf{Q}}) \tilde{\Phi}_{\nu\mathbf{Q}}(\mathbf{k}) - \sum_\mathbf{q} W_\mathbf{q} \tilde{\Phi}_{\nu\mathbf{Q}}(\mathbf{k+q}) = E_{\nu\mathbf{Q}} \tilde{\Phi}_{\nu\mathbf{Q}}(\mathbf{k}).
\end{eqnarray}
Within the vicinity of minima and maxima of valence and conduction band, we approximate the dispersions quadratically, which allows us to separate relative and center of mass motion. When $\mathbf{K}_c$ denotes the conduction band valley and $\mathbf{K}_v$ the valence band valley, we find $\tilde{\Phi}_{\nu\mathbf{Q}}(\mathbf{k})=\tilde{\Phi}_{\nu}(\mathbf{k})=\tilde{\Psi}_\nu(\mathbf{k}-\alpha \mathbf{K}_v-\beta \mathbf{K}_c)$, with $\tilde{\Psi}$ obeying the effective electron-hole Schroedinger equation,

\begin{eqnarray} \label{eq:wannier2}
\frac{\hbar^2 k^2}{2 m_{\text{r}}} \tilde{\Psi}_\nu(\mathbf{k}) - \sum_\mathbf{q} W_\mathbf{q} \tilde{\Psi}_\nu(\mathbf{k+q}) = E^\text{bind}_\nu\tilde{\Psi}_\nu(\mathbf{k}),
\end{eqnarray}

where $m_\text{r}=(m_cm_v)/(m_c+m_v)$ is the reduced exciton mass for the corresponding valley masses of electrons ($m_c$) and holes ($m_v$). Furthermore, the parabolic approximation yields $E_{\nu\mathbf{Q}}=E^\text{bind}_\nu+\hbar^2(\mathbf{Q}-[\mathbf{K_c}-\mathbf{K_v}])^2/(2[m_c+m_v])+\varepsilon_{c\mathbf{K_c}}-\varepsilon_{v\mathbf{K_v}}$. Note, that exciton wavefunctions with different valley configurations are centered at different momenta. Therefore, the exciton form factor for scattering from $\nu=(n,K-K)$ to $\mu=(m,K-\Lambda)$ reads

\begin{eqnarray} \label{eq:formfac2}
\mathcal{F}_{\nu\mu}(\alpha_\mu\mathbf{q})=\sum_{\mathbf{k}} \tilde{\Phi}_{\nu}^{\ast}(\mathbf{k}) \tilde{\Phi}_\mu(\mathbf{k+\alpha_\mu q}) =\sum_{\mathbf{k}} \tilde{\Psi}_{n}^{\ast}(\mathbf{k}) \tilde{\Psi}_m(\mathbf{k}+\alpha_\mu [\mathbf{q}-(\Lambda-K)])
\end{eqnarray}

Hence, the wavefunction overlap in Eq. \eqref{eq:formfac2} gets maximized for a momentum transfer $\mathbf{q}=\Lambda-K$ conecting K and $\Lambda$ valley, while the overlap for intravalley scattering decreases with growing q. Furthermore, to account for the complex phase of the electron-phonon coupling, we discard mixed terms in the calculation of the exciton-phonon coupling, i.e.

\begin{eqnarray} \label{eq:average}
\lvert G^{\nu \mu}_{\lambda\mathbf{q}}\rvert^2 \approx \sum_{\eta=e,h} \lvert g^{\eta}_{\lambda \mathbf{q}} \mathcal{F}_{\nu\mu}(\mathbf{q}_\eta)\rvert^2,
\end{eqnarray}

assuming similar weights for polar and non-polar coupling mechanism \cite{kaasbjerg2013acoustic} (e.g. optical deformation potential vs. Froehlich interaction) .

\newpage
\section{Experimental Procedure and Linewidth Extraction}

The studied hBN-encapsulated WSe$_2$ sample was obtained by mechanical exfoliation of hBN and WSe$_2$ flakes onto polydimethylsiloxane (PDMS) film and subsequent stacking via stamp transfer\,\cite{Castellanos-Gomez2014} onto SiO$_2$/Si substrate. 
First, thin hBN flakes (provided by T. Taniguchi and K. Watanabe, NIMS) were stamped onto a $100^\circ \text{C}$ preheated $\unit[295]{nm}$ thick SiO$_2$/Si substrate at ambient conditions. 
Then a WSe$_2$ monolayer was stamped on top of the hBN layer, followed by placing an additional hBN layer on top of the structure at $70^\circ \text{C}$ substrate temperature and ambient conditions.
The sample was annealed in high vacuum at $150^\circ \text{C}$ for 4-5 hours after each individual stamping process. 
The sample was scanned to find large, homogeneous areas of several micrometers with narrow exciton resonances in both light emission and reflectance, indicating successful transfer and good interlayer coupling.

Reflectance measurements were performed using a spectrally broadband tungsten-halogen lamp for illumination.
The reflected signals were collected both on the sample ($R_{s}$) and on the SiO$_2$/Si substrate reference (R$_{r}$), spectrally dispersed in a grating spectrometer, and detected by a cooled CCD camera. 
The sample was placed in an optical microscopy cryostat cooled by liquid helium. The heat sink temperature was tuned between 4 and 300\,K, giving the system sufficient time to reach equilibrium between individual measurements.
The lattice temperature of the sample was independently confirmed by the relative energy shift of the exciton resonance.  

\begin{figure}[b]
    \centering
        \includegraphics[width=16.5 cm]{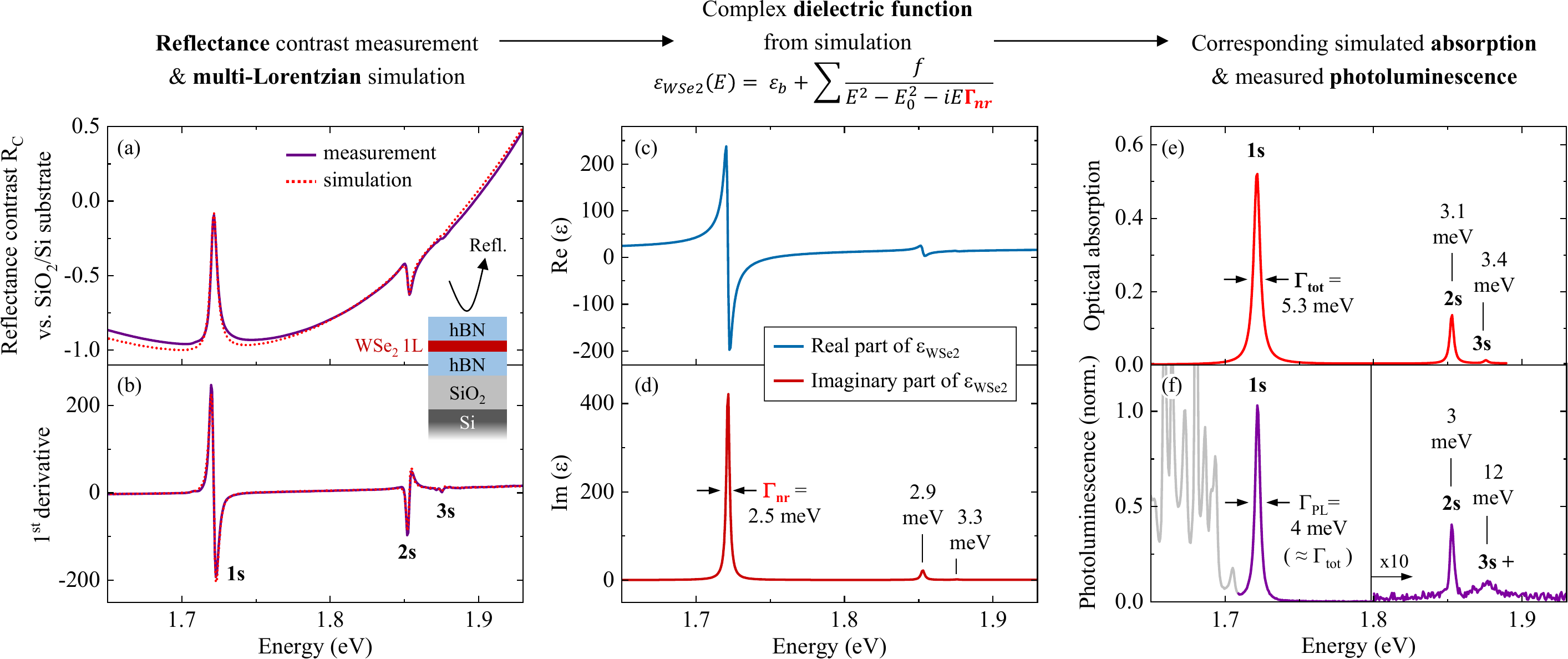}
        \caption{(a) Reflectance contrast of the hBN-encapsulated WSe$_2$ monolayer at 4\,K together with the simulated spectrum from the multi-Lorentzian model.
                (b) First-order derivative of the measured and simulated reflectance contrast.
                (c), (d) Real and imaginary parts of the parameterized dielectric function used in the simulation. 
                In the imaginary part, non-radiative linewidths $\Gamma_{nr}$ of the ground and the first two excited stated exciton transitions are indicated.
                (e) Corresponding simulated optical absorption spectrum of the $n=1, 2, 3$  exciton resonances.
                Total linewidths are indicated by $\Gamma_{tot}$.
                (f) Measured continuous-wave photoluminescence spectrum of the same sample at 4\,K.
        }
\label{figSI-exp1}
\end{figure}

For the analysis, the acquired reflectance signals are presented in terms of reflectance contrast, defined as $R_{C} = (R_{s}-R_{r})/(R_{r}-R_{bg})$, where $R_{bg}$ denotes the background signal without illumination. 
$R_{C}$ thus corresponds to the relative change in the reflectance of the sample with respect to the bare SiO$_2$/Si substrate. 
A representative reflectance contrast spectrum of the WSe$_2$ monolayer at the temperature of 4\,K is shown in Fig.\,\ref{figSI-exp1}\,(a). 
The corresponding first derivative of $R_{C}$ with respect to the photon energy is plotted in Fig.\,\ref{figSI-exp1}\,(b).
In the studied spectral range, the optical response of the sample is dominated by the A exciton ground state resonance (1s) at 1.721\,eV, first excited state transition (2s) at 1.853\,eV and a weak feature of the second excited state (3s) at 1.876\,eV.
The absence of pronounced features below the 1s resonance indicate negligible free charge carrier densities.

To extract the exciton peak parameters, the energy-dependent dielectric function of the WSe$_2$ monolayer $\varepsilon(E)$ is parameterized with multiple Lorentzian resonances:
\begin{equation}
\varepsilon(E)=\varepsilon_b+\sum_{j=1}^{N}\frac{f_j}{E_{j}^2-E^2-iE\Gamma_{nr, j}},
\end{equation}
where $f_j$, $E_j$, and $\Gamma_{nr, j}$ represent the oscillator strength, peak energy, and the purely \textit{non-radiative} damping of the resonance with the index $j$, respectively.
The linewidth is defined as full-width-half-maximum of the respective peak in the imaginary part.
Only a small number $N$ of the resonances is included in the simulation. 
In the spectral range of interest, only those clearly visible in the measured spectra are considered, i.e., the exciton 1s, 2s, and 3s states.
The reflectance contrast is then computed using a transfer-matrix formalism\,\cite{Byrnes2012} taking into account multi-layer interference effects due to the presence of the hBN layers and the SiO$_2$/Si substrate . 
For the studied structure, the thickness of the SiO$2$ layer was set to 296\,nm and the thickness of the top and bottom hBN layers - to 16.4\,nm (using refractive index of 2.2) to obtain the measured overall spectral shape of the reflectance contrast.
The exciton peak parameters are then adjusted to match the measured first order derivative.
The simulated spectra are presented alongside experimental data in Figs.\,\ref{figSI-exp1}\,(a) and (b), exhibiting good agreement and allowing for a reasonable extraction of the exciton peak parameters.

Real and imaginary parts of the dielectric function corresponding to the simulated response are presented in Figs.\,\ref{figSI-exp1}\,(c) and (d), respectively. 
We note, that the width of the resonances in the imaginary part is determined only by the \textit{non-radiative} broadening $\Gamma_{nr}$.
In contrast to that, \textit{total} linewidths of the same peaks in the optical absorption include additional broadening due to the radiative coupling, i.e., finite oscillator strength.
The absorption spectrum computed from the same dielectric function is shown in Fig.\,\ref{figSI-exp1}\,(e). 
For the 1s resonance, in particular, the total linewidth $\Gamma_{tot}$ of the absorption peak is extracted to be 5.3\,meV compared to the purely non-radiative broadening $\Gamma_{nr}$ of 2.5\,meV. 
Total linewidths are used for comparison with theory throughout the paper.

Here, we note that the resonance linewidths in the emission spectra should also correspond to the total linewidths from absorption, i.e., include both radiative and non-radiative contributions.
For comparison, a representative photoluminescence (PL) spectrum obtained at roughly the same position on the WSe$_2$ sample as the reflectance measurements is shown in Fig.\,\ref{figSI-exp1}\,(f).
A continuous-wave laser emitting at 532\,nm was used for the excitation with a power of 10\,$\mu$W focused to a spot of about 1\,$\mu$m. 
The emission from the exciton ground and excited states is highlighted in the data.
The signals associated with more complex exciton states below the 1s resonance (trions, biexcitons, localized states etc.) are shown in gray.
The linewidth of the exciton resonance is found to be in the range of 4-5\,meV in PL, close to the values obtained in the absorption spectra.

\section{Temperature Dependent Spectra}

\begin{figure}[b]
    \centering
        \includegraphics[width=7.5 cm]{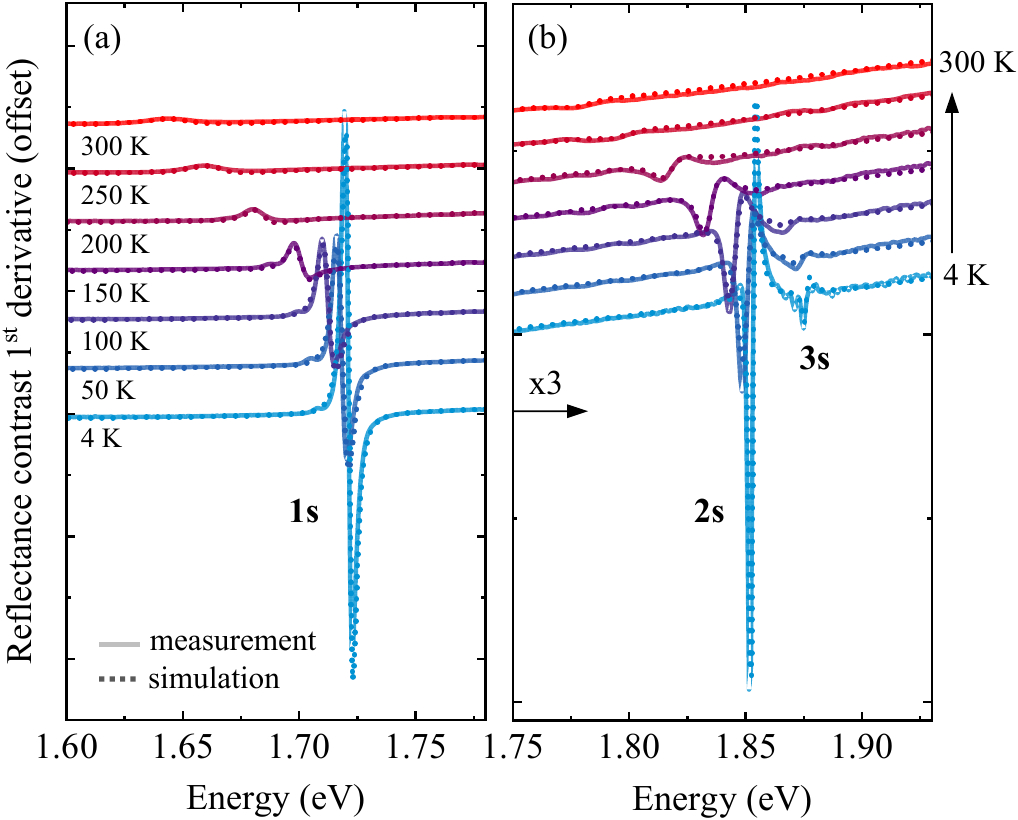}
        \caption{(a) Measured reflectance contrast derivatives of the hBN-encapsulated WSe$_2$ sample for temperatures from 4 to 300\,K together with the simulation results.
        The spectra are presented in the spectral range of the 1s ground state exciton resonance.
        (b) Same as (a) in the spectral range of excited state resonances.
        The data are vertically offset for clarity.
        }
\label{figSI-exp2}
\end{figure}

Experimentally measured reflectance contrast spectra are presented as first derivatives in the Fig.\,\ref{figSI-exp2}\,(a) and (b) in the spectral range of ground and excited state resonances, respectively.
The simulation results from the multi-Lorentzian peak analysis, discussed in the previous section, are plotted alongside the measured data.
Three exciton states, 1s, 2s, and 3s, are observed at 4\,K.
The 2s state is detected up to room temperature and the 3s state is resolved at 100\,K and below.
As the temperature increases, the exciton resonances shift to lower energies in good agreement with the literature results on as-exfoliated samples\,\cite{Arora2015}.
We note that the shifts of the ground and excited states are very similar, i.e., the shift of the 2s state closely follows that of the 1s resonance with small deviations on the order of several meV towards room temperature.
The energy shifts are accompanied by pronounced spectral broadening due to exciton-phonon scattering, analyzed and discussed in the main manuscript in detail.

\end{document}